\documentclass[letter,structabstract]{aa} % for the letters 

\usepackage{graphicx}
\usepackage{txfonts}
\usepackage{natbib}

\bibpunct{(}{)}{;}{a}{}{,}

\begin{document}
   \title{The kinematic of HST-1 in the jet of M87}

\author{
M.~Giroletti\inst{1} \and 
K.~Hada\inst{2,3} \and 
G.~Giovannini\inst{1,4} \and 
C.~Casadio\inst{5} \and 
M.~Beilicke\inst{6} \and 
A.~Cesarini\inst{7} \and 
C.C.~Cheung\inst{8} \and
A.~Doi\inst{9} \and 
H.~Krawczynski\inst{6} \and 
M.~Kino\inst{3} \and 
N.P.~Lee\inst{10} \and
H. Nagai\inst{3}
}

\institute{
  INAF Istituto di Radioastronomia, 40129 Bologna, Italy \email{giroletti@ira.inaf.it} \and
  The Graduate University for Advanced Studies (SOKENDAI), 2-21-1 Osawa, Mitaka, Tokyo 181-8588, Japan \and
  National Astronomical Observatory of Japan, 2-21-1 Osawa, Mitaka, Tokyo, 181-8588, Japan \and
  Dipartimento di Astronomia, via Ranzani 1, 40127 Bologna, Italy \and
  Instituto de Astrofisica de Andalucia, CSIC, Apartado 3004, 18080, Spain \and
  Department of Physics, Washington University, St. Louis, MO 63130, USA \and
  School of Physics, National University of Ireland, University Road, Galway, Republic of Ireland \and
National Research Council Research Associate, National Academy of
Sciences, Washington, DC 20001, resident at Naval Research Laboratory,
Washington DC 20375, USA \and
  Institute of Space and Astronautical Science, JAXA, 3-1-1 Yoshinodai, Sagamihara, Kanagawa 229-8510, Japan \and
Smithsonian Astrophysical Observatory, 60 Garden St., Cambridge MA 02138, USA
             }

\date{Received ; accepted }

\abstract
  {}
  {We aim to constrain the structural variations within the HST-1 region
    downstream of the radio jet of M87, in general as well as in connection to the episodes of
    activity at very high energy (VHE).}
  {We analyzed and compared 26 VLBI observations of the M87 jet, obtained between
    2006 and 2011 with the Very Long Baseline Array (VLBA) at 1.7 GHz and the
    European VLBI Network (EVN) at 5 GHz.}
  {HST-1 is detected at all epochs; we model-fitted its complex structure with two
    or more components, the two outermost of which display a significant proper
    motion with a superluminal velocity around $\sim 4\,c$. The motion of a third feature that is
    detected upstream is more difficult to characterize. The
    overall position angle of HST-1 has changed during the time of our
    observations from $-65^\circ$ to $-90^\circ$, while the structure has moved
    by over 80 mas downstream. Our results on the component evolution
    suggest that structural changes at the upstream edge of HST-1 can be
    related to the VHE events.}
  {}


\maketitle
%
%________________________________________________________________

\section{Introduction}

The debate about the location and the mechanisms for the production of MeV/GeV,
and very high energy (VHE) gamma rays in Active Galactic Nuclei (AGN) jets is very lively in the era of
the {\it Fermi} satellite and the new generation Cherenkov telescopes (VERITAS,
MAGIC, HESS). M87 is a particularly suitable laboratory for a detailed study of the
properties of jets, given its proximity ($D=16$ Mpc), the massive black hole
\citep[$M_\mathrm{BH}\sim 6.4 \times 10^9 M_\odot$,][corresponding to a scale
  of 1 mas $\sim 150 R_S$]{Gebhardt2009}, and its conspicuous emission at all
wavelengths.  M87 shows a prominent radio, optical, and X-ray jet,
characterized by many substructures and knots from sub-parsec to kiloparsec
scale.  VLBI observations of the inner jet show a well-resolved,
edge-brightened structure, which starts with a wide opening angle
\citep{Junor1999,Krichbaum2005,Ly2007} and then experiences a strong
collimation, with an opening angle smaller than 10$^\circ$ \citep{Kovalev2007}.

At about 0.8--0.9 arcseconds from the core, the jet suddenly re-brightens. This
feature was first discussed in the optical by \citet{Biretta1999}, who named it
HST-1 and showed that it was moving at $0.84\,c$; it also appeared to emit
superluminal optical features with velocity $\sim 6\,c$. Superluminal
components within HST-1 were later found with much finer angular resolution
thanks to VLBI observations at 1.7 GHz by \citet{Cheung2007}. In addition to
presenting this hallmark of blazar activity, HST-1 underwent a dramatic
brightening in radio, optical, and X-rays during 2003-2006, becoming even
brighter than the nucleus in X-rays \citep{Harris2006}. These facts together
led \citet{Cheung2007} to propose that the flaring activity registered at VHE
in 2005 \citep{Aharonian2006} originated within HST-1.  On the other hand,
the VHE variability on time scales of days seemed to require a much more
compact emission region, suggesting the nucleus of M87 itself as a likely site
of TeV $\gamma$-ray production \citep{Aharonian2006}, involving, e.g., the
BH magnetosphere \citep{Neronov2007} or an interaction between a fast
jet spine and a slower sheath \citep{Tavecchio2008}. Indeed, a second VHE flare
was observed in 2008, simultaneously to a strong increase of the 43 GHz flux
density of the core, while HST-1 was in a low state \citep{Acciari2009}.

In this context, we started at the end of 2009 a program to monitor M87 at 5GHz
with the European VLBI Network in real-time mode \citep[e-EVN,][]{Szomoru2008}
during the season of VHE observations in 2009/2010. The chosen array
configuration provides a suitable combination of resolution (down to just above
1 mas), sensitivity (a few $\times 0.1$ mJy\,beam$^{-1}$), and field of view
(several arcseconds), which permits a detailed study of the behavior of both the
core and HST-1. Indeed, VHE activity was observed twice in 2010, very near in
time to one of our e-EVN observations \citep{Giroletti2010}.  While a detailed
multifrequency study of M87 during the VHE event is presented in a dedicated
paper \citep{Abramowski2012}, in this letter we focus on the kinematic
properties of HST-1, extending the dataset of the EVN observations with
archival VLBA data at 1.7 GHz, dating back to 2006. The new and archival
observations are presented in Sect.~\ref{s.observations}; in
Sect.~\ref{s.results}, we describe the main results, and we discuss the
implication and summarize the main conclusions in Sect.~\ref{s.discussion}.
Throughout the paper, we assume for M87 a distance of 16 Mpc \citep{Tonry2001},
corresponding to a scale of 1\,mas\,=\,0.078\,pc; proper motion of 1 mas/yr
corresponds to an apparent speed of $0.25\,c$. The spectral index $\alpha$ is
defined such that $S(\nu)\sim\nu^{-\alpha}$.

\section{Observations and data reduction\label{s.observations}}

   \begin{figure}
   \centering
   \includegraphics[width=0.92\columnwidth]{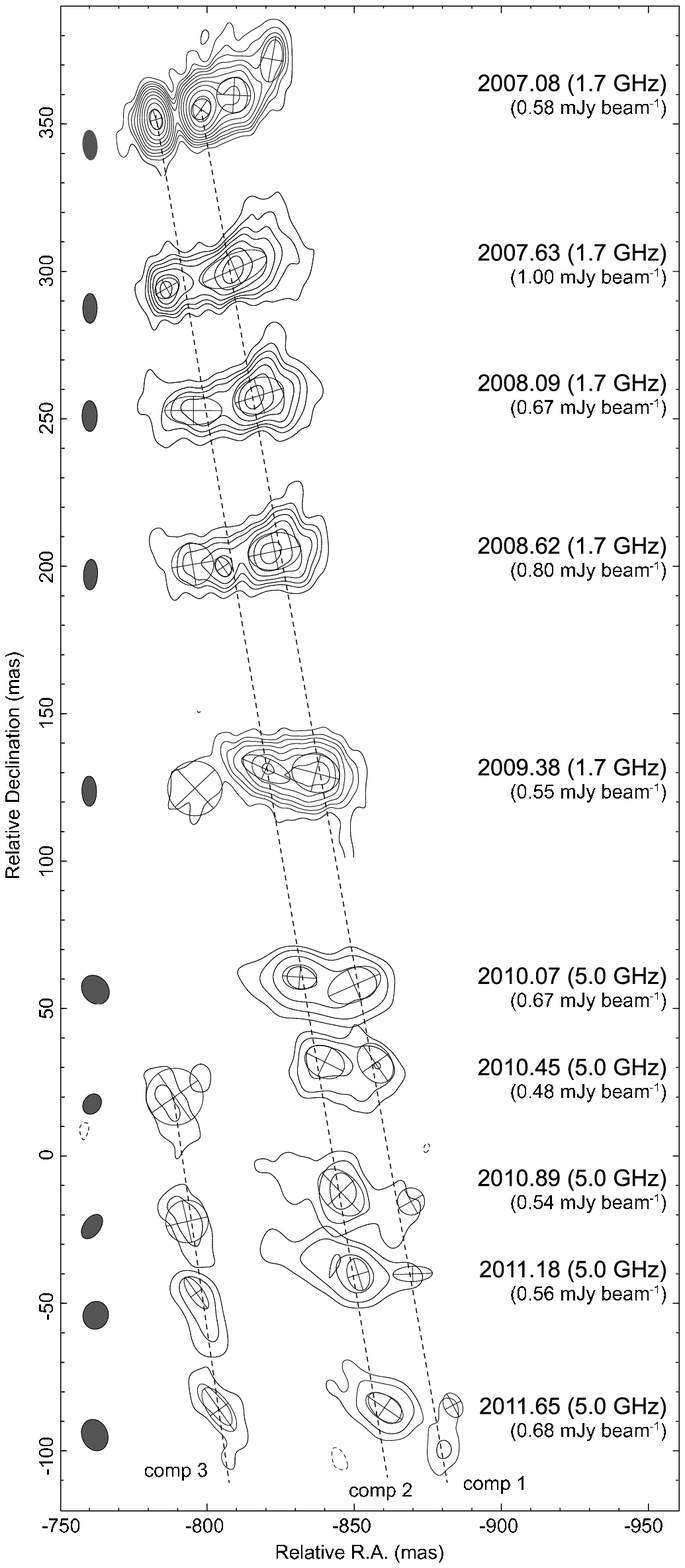}%scale=0.65
      \caption{Set of HST-1 images. For each epoch, we give the epoch, frequency, lowest contour to the right, and the restoring beam to the left of the respective contour plot. Contours are traced at $(-1, 1, 2, 4, \dots)$ multiples of the given value. The model-fit components are overlaid as ellipses with crosses. The contour plots are spaced vertically proportionally to the time interval between the relative epochs. The axes represent the relative (RA, Dec) coordinates from the core for the first image.
              }
         \label{f.images}
   \end{figure}

\subsection{e-EVN data}

We observed M87 with the e-EVN at 13 epochs between June 2009 and October 2011.
The observing dates are given in Table~\ref{t.log}. Typically, observations lasted
4--8 hours, and the longest baselines were achieved from European stations to Shanghai
and/or Arecibo. Some observations are somewhat limited in quality,
because they were obtained as targets of opportunity, or from calibration
observations. These observations are noted in Table~\ref{t.log}; overall,
the data quality is adequate to warrant good signal-to-noise detections of the
source structure.

For all observations, the frequency setup was centered at 5.013 GHz and divided
into eight sub-bands separated by 16 MHz each for an aggregate bit rate of 1\,Gbps.
The data were correlated in real time at JIVE, except for the first 
observation, which was disk-recorded; automated data flagging and initial
amplitude and phase calibration were also carried out at JIVE using dedicated
pipeline scripts. The data were finally averaged in frequency within each IF, but
individual IFs were kept separate to avoid bandwidth smearing.  Similarly, the
data were time-averaged only to 8\,s to avoid time smearing. 

We produced final images after several cycles of phase and amplitude
self-calibration. We applied Gaussian tapers to the visibility data to maximize sensitivity to the faint and extended emission in HST-1, resulting in a beam in the range $5-10$ mas and rms noise of $0.1-0.3$ mJy/beam.

\subsection{VLBA data}

M87 has been observed with the VLBA at 1.7 GHz 12 times between 2006
November and 2009 August. Each of these datasets has an on-source time of $\sim
6$ hours, with a total bandwidth of 32 MHz. We also added two observations in
2010 April, taken at 2.3 GHz, which were part of a multi-frequency VLBA
experiments aimed at studying the opacity in the core of M87
\citep{Hada2011}. Each session has a total on-source time of $\sim$15 minutes
with a total bandwidth of 32~MHz. Observing dates are summarized in
Table~\ref{t.log}.

The initial data calibration was performed in NRAO AIPS based on the standard
reduction procedures. The data were averaged at short intervals (5 seconds in
time and 1 MHz in frequency) to minimize smearing effects. The data of two
multi-frequency experiments, which were separated by only 10 days, were
combined to improve the image quality.  All images were made with
iterative phase and amplitude self-calibration.  The resulting image rms noises
in the HST-1 region are 0.1$-$0.3 mJy/beam. The naturally weighted images provide
a typical beam size of about 10.5 $\times$ 5.5 mas elongated in declination.

Overall, the final resolution and sensitivity of the VLBA and EVN data sets are well-matched.
The presence of short and sensitive baselines in the EVN compensates for the spectral and resolution effects caused by the higher observing frequency.

\begin{table*}
\begin{minipage}[t]{\textwidth}
\caption{Log of observations.
}
\label{t.log}
\centering
\renewcommand{\footnoterule}{}  % to avoid a line before footnotes
\begin{tabular}{lll}
\hline \hline
Array & Observing & Observation date \\
      & Frequency &                  \\
\hline
VLBA  & 1.7 GHz   & 2006.86, 2007.08, 2007.41, 2007.63, 2007.95, 2008.09, 2008.40, 2008.62, 2008.91, 2009.14, 2009.38, 2009.64, 2010.28$^a$ \\
EVN   & 5 GHz     & 2009.45$^b$, 2009.88, 2010.07, 2010.11, 2010.18$^c$, 2010.24, 2010.38, 2010.45, 2010.89, 2011.18, 2011.28, 2011.65, 2011.79 \\
\hline
\hline
\end{tabular}
\end{minipage}
Note: $^a$ 2.3 GHz observations, averaged date for the 2010.27 and 2010.29; $^b$ calibrator observations; $^c$ ToO observation, only six stations and 128 Mbps available for self-calibration.
\end{table*}

\section{Results\label{s.results}}

   \begin{figure}
   \centering
   \includegraphics[width=0.97\columnwidth,clip]{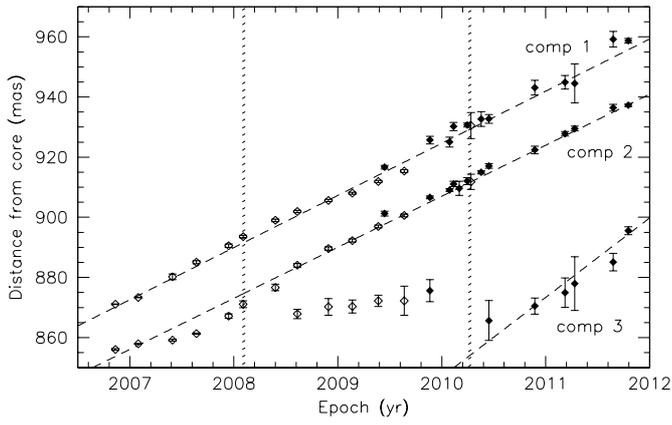}
      \caption{Distance of the HST-1 sub-components from the core versus time. Empty and filled symbols show 1.7 (VLBA) and 5 GHz (EVN) data, respectively. The dashed lines show least-squares linear fits, while the vertical stripes represent the epochs of the 2008 and 2010 VHE flares.
              }
         \label{f.motion}
   \end{figure}

\begin{table}
\caption{Component motion results}
\label{t.motion}
\centering
\begin{tabular}{c c c c}          % centered columns (4 columns)
\hline\hline                        % inserts double horizontal lines
Component & Interval & $\Delta r$ & $\beta_{\rm app}$ \\
Name      & (yy)     & (mas)      & $(v/c)$           \\
\hline                                   % inserts single horizontal line
 1 & 2006.86-2011.80 & 88.1 & $4.17\pm0.07$ \\
 2 & 2006.86-2011.80 & 81.2 & $4.08\pm0.08$ \\
 3 & 2010.45-2011.80 & 29.8 & $6.4\pm0.8$ \\
\hline
\end{tabular}
\end{table}

We detect significant flux density in the HST-1 region at all epochs, in
addition to the bright core and the inner jet. The details of the imaging
somewhat change depending on the observing frequency, the observation epoch,
the $(u,v)$-plane coverage, and the adopted weighting scheme. In general, the
flux density in the EVN images ($\langle S_{5}\rangle\sim23$~mJy) is lower than
in the VLBA ones ($\langle S_{1.7}\rangle\sim90$~mJy), resulting in a non
simultaneous spectral index of $\alpha=1.2$.

The HST-1 region extends for over 50 mas and is resolved in complex
substructures. The overall position angle and the location of the individual
substructures evolve with time. We model-fitted the visibility data in Difmap
for all epochs, adopting elliptical Gaussian components to describe the
emission from HST-1. In Fig.~\ref{f.images}, we show ten contour images of the
HST-1 region, overlaid with model-fit components. Two or three components are
usually required to describe the visibility data. In Fig.~\ref{f.motion} we
plot the separation of the components from the core as a function of time. The
overall uncertainty on the position at each epoch is estimated taking into
account the component size and the signal-to-noise ratio. There is an
additional uncertainty related to opacity effects, since the core position is
not the same at the different observing frequencies. However, we estimate that
any possible core-shift is significantly smaller than our estimated
uncertainty on the basis of the results reported by \citet{Hada2011} and of
the large overall displacement of each individual component.

Thanks to the many observations and the good accuracy in fitting the
structure, we can reliably track the components between the various epochs. In
particular, we are confident of the identification of the two main components,
which we label as component 1 and 2, of which 1 is the outermost. The size of each
component varies and additional components are present at some
epochs. In particular, component 2 becomes quite extended in early 2008 and
eventually splits into two components from 2008.62. After this split, the
upstream subcomponent remains more or less stationary and gradually becomes
fainter. When the 5 GHz observations start, there is little evidence of this component. However, starting from 2010.45, a new inner
component is again required to fit the 5 GHz data. This component appears
consistently thereafter and we name it component 3.

In Table \ref{t.motion}, we summarize the displacement and velocity of each
component during the campaign. Overall, components 1 and 2 have moved by very
similar distances ($\Delta r_1 = 88.1$ mas, $\Delta r_2 = 81.2$ mas),
corresponding to apparently superluminal velocities around $\beta_{\rm
  app}=4.1$. Thanks to the many observations, the uncertainty on
this superluminal value can be constrained down to as small as a few percent.
The faint substructure (identified as component 3) visible only in the last
epochs at $\sim 860-890$ mas from the core is also moving superluminally, but
the uncertainty is larger because of the shorter time range.  The
identification of this component with the inner, slower feature resulting from
the splitting of component 2 in 2008.62 is not straightforward.

   \begin{figure}
   \centering
   \includegraphics[clip,width=0.94\columnwidth]{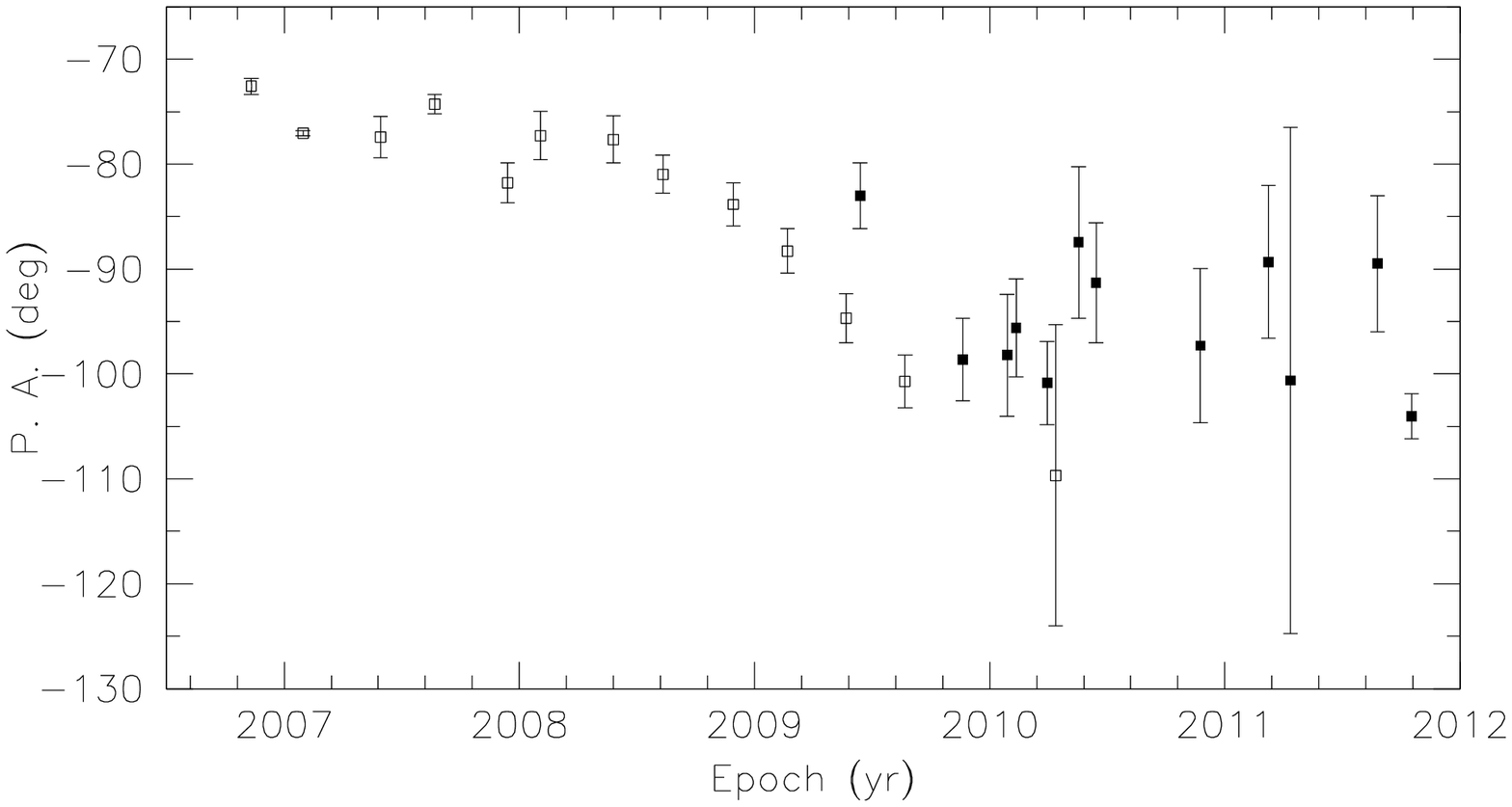}
      \caption{Orientation of the vector connecting the two main components in HST-1 as a function of time.
              }
         \label{f.angle}
   \end{figure}

In Fig.~\ref{f.angle}, we show the change of PA of the main emission in the
HST-1 region, represented by the orientation of the vector that connects
components 1 and 2. At the early epochs, HST-1 is oriented similarly to the
main jet of M87 ($\sim -65^\circ$). In the following epochs, the position angle
progressively changes to $-90^\circ$ and finally to $\sim -100^\circ$ as the
components move outward. In the final observations, the PA distribution
presents more scatter because of the lower S/N in the data.

\section{Discussion\label{s.discussion}}

The observations presented in this letter clearly demonstrate that HST-1 is
resolved in complex substructures. Two main components can be reliably
identified across epochs and they are found to move with apparently
superluminal velocity ($\sim 4c$). From the observed speed, we can infer a
range of the possible intrinsic jet velocity, assuming that the pattern and bulk
velocity are the same.

The orientation angle in M87 has been widely debated in the last years. Large
Doppler factors are required by the high-energy emission and the radio
properties \citep[jet brightness, superluminal motions downstream the jet,
  e.g.][]{Acciari2009}. Therefore, the jet has to be closely aligned with the
line of sight. However, the evidence of a limb-brightened structure
\citep{Ly2007,Kovalev2007} suggests a relatively wide jet orientation angle
with respect to the line of sight \citep{Giroletti2004}.  Assuming a jet
orientation angle in the range $15^\circ < \theta < 25^\circ$, the measured
apparent velocity of $4\,c$ corresponds to an intrinsic velocity $0.97\,c < v <
0.99\,c$, which in turn implies a Doppler factor and a Lorentz factor for this
structure between $\delta_{\rm HST1}=1.5$ and $\Gamma_{\rm HST1}=6.5$ (for
$\theta=25^\circ$), and $\delta_{\rm HST1}=3.9$ and $\Gamma_{\rm HST1}=4.1$
(for $\theta=15^\circ$), respectively.  This result agrees with the
synchrotron model for the X-ray emission discussed by \citet{Marshall2002} and
\citet{Harris2003}.  By contrast, different values of the apparent velocity
have been reported by other authors in other parts of the jet and/or using data
obtained in different epochs
\citep{Biretta1999,Cheung2007,Kovalev2007,Ly2007,Asada2011}, suggesting that
the velocity structure in this jet is quite complex.

At first sight, our result does not support the identification of HST-1 as a
standing-shock structure, given the displacements of $>80$ mas found for
components 1 and 2 over $\sim 5$ years. Moreover, no prominent stationary
components were  consistently detected in the 26 observations considered in
this work.
However, the components change in size over time, and in particular component 2
splits into substructures after 2008.62, suggesting that other features exist
in HST-1 and interact with the brightest knots. Those components may be
underlying, standing or very slowly moving regions, too faint to be detected
separately but contributing to the total emission when brighter components are
nearby. In particular, this could be the case for the D component of
\citet{Cheung2007}, which only becomes visible when a new feature (like our
component 3) is ejected/created within HST-1.
The emergence of the new superluminal component 3, upstream of 1 and 2,
accompanied by a brightness decrease of 1 and 2, could eventually shift back
the centroid of the arcsecond scale emission of HST-1, which would then appear
to be more stationary than individual substructures. This behavior would then
support the scenario in which HST-1 is a stationary reconfinement shock
structure, possibly associated to a jet interaction with a gaseous condensation
of the hot interstellar medium \citep{Stawarz2006}.

Based on the short variability time scales of the VHE events and on the
recurrent simultaneous X-rays brightening of the core, the viability of HST-1
as the site of emission of VHE radiation has recently been  severely questioned 
\citep{Abramowski2012}. We show the epochs of the two latest VHE flares from
M87 with dotted lines in Fig.~\ref{f.motion}. Both events are followed by
structural changes and rebrightening of the upstream edge of HST-1, suggesting
that the origin of the VHE activity could indeed be related to the HST-1
region. A similar connection was also put forward for the 2005 VHE event: by
considering a subset of the present dataset and archival VLA observations,
\citet{Giovannini2011} noted a change in the proper motion velocity in HST-1 at
the epoch $\sim$ 2005.5, coincident with the TeV $\gamma$-ray activity and the
maximum radio and X-ray flux density of the feature.

Lastly, we can discuss the propertis of the magnetic field parallel to the
shock front in HST-1, based on the field strength in the unshocked jet and the
magnetic flux conservation. By fitting a single-zone synchrotron self-Compton
model, \citet{Abdo2009} estimated a magnetic field strength at the core $B_{\rm
  core} \sim 0.055\,{\rm G}$. The field strength in the unshocked jet upstream
of HST-1 can be estimated as $B_{\rm jet}=(r_{\rm HST1}/r_{\rm core})^{-1}
B_{\rm core} \approx 0.33\times 10^{-3} (B_{\rm core}/0.1~{\rm G})~{\rm G}$,
where we assume $Br=\mathrm{const}$ and the jet radius in HST-1 as $r_{\rm
  jet}\approx r_{\rm HST1}\approx 15~{\rm mas} \approx 4\times 10^{18}~{\rm
  cm}$, based on the images in Fig.~\ref{f.images}.
Next, let us assume that the shock dissipation occurs at HST-1.  Then we can
estimate the field strength downstream of the unshocked jet by the shock jump
conditions.
A typical strength of the compressed magnetic field at HST-1 ($B_{\rm HST1}$)
can be estimated by $B_{\rm HST1}\approx \Gamma_{\rm jet} B_{\rm jet} \approx
2.3 (\Gamma_{\rm jet}/7) (B_{\rm core}/0.1~{\rm G})~{\rm mG}$, where $\Gamma_{\rm
  jet}$ is the bulk Lorentz factor of the unshocked jet at the HST-1-upstream,
which satisfies $\Gamma_{\rm jet}>\Gamma_{\rm HST1}$.
Although roughly comparable to the independently estimated value of $B_{\rm
  HST1}\delta_{\rm HST1}^{1/3}=1.1~{\rm mG}$ by using synchrotron cooling time
at X-ray band \citep{Harris2003,Harris2009}, the $B_{\rm HST1}$ estimated here
tends to be higher by a factor of several; clarifying the reasons of the
discrepancy remain to be worked out.
% A slightly larger $r_{\rm HST1}$ due to very faint surrounding structure
%could explain the discrepancy.

Additional discussions on the magnetic field structure, as well as the light
curve and trajectories of the individual component will therefore be treated in
upcoming publications, also in the light of new planned VLBI
observations. Indeed, our original aim was to address two main questions: (1)
Where are the high energy gamma-rays produced in M87: the core or HST-1? (2) Is
HST-1 stationary or moving?  The answer to both questions may not be too
simple!

\begin{acknowledgements}
We thank D.\,E.\,Harris for reading the manuscript.  We acknowledge a
contribution from the Italian Foreign Affair Minister under the bilateral
scientific collaboration between Italy and Japan.  e-VLBI research
infrastructure in Europe is supported by the European Union's Seventh Framework
Programme (FP7/2007-2013) under grant agreement no.\ RI-261525 NEXPReS.  The
European VLBI Network is a joint facility of European, Chinese, South African
and other radio astronomy institutes funded by their national research
councils. The National Radio Astronomy Observatory is a facility of the
National Science Foundation operated under cooperative agreement by AUI.
\end{acknowledgements}

%\vspace{-0.4cm}

\end{document}